\documentstyle[12pt,twoside]{article}
\def\greaterthansquiggle{\raise.3ex\hbox{$>$\kern-
.75em\lower1ex\hbox{$\sim$}}}
\def\lessthansquiggle{\raise.3ex\hbox{$<$\kern-
.75em\lower1ex\hbox{$\sim$}}}
\newcommand{\beq}{\begin{equation}}
\newcommand{\eeq}{\end{equation}}
\newcommand{\beqa}{\begin{eqnarray}}
\newcommand{\eeqa}{\end{eqnarray}}
\newcommand{\beqan}{\begin{eqnarray*}}
\newcommand{\eeqan}{\end{eqnarray*}}
\newcommand{\ba}{\begin{array}}
\newcommand{\ea}{\end{array}}
\newcommand{\no}{\nonumber}

\newcommand{\Un}{\underline}

\newcommand{\ra}{\rightarrow}

\newcommand{\vp}{\varphi}

\newcommand{\wt}{\widetilde}

\newcommand{\A}{{\cal A}}

\newcommand{\E}{{\cal E}}

\newcommand{\G}{{\cal G}}

\def\nz{\ifmmode {I\hskip -3pt N} \else {\hbox {$I\hskip -3pt N$}}\fi}
\def\zz{\ifmmode {Z\hskip -4.8pt Z} \else
       {\hbox {$Z\hskip -4.8pt Z$}}\fi}
\def\qz{\ifmmode {Q\hskip -5.0pt\vrule height6.0pt depth 0pt
       \hskip 6pt} \else {\hbox
       {$Q\hskip -5.0pt\vrule height6.0pt depth 0pt\hskip 6pt$}}\fi}
\def\rz{\ifmmode {I\hskip -3pt R} \else {\hbox {$I\hskip -3pt R$}}\fi}
\def\cz{\ifmmode {C\hskip -4.8pt\vrule height5.8pt\hskip 6.3pt} \else
       {\hbox {$C\hskip -4.8pt\vrule height5.8pt\hskip 6.3pt$}}\fi}

\def\au{{\setbox0=\hbox{\lower1.36775ex%
\hbox{''}\kern-.05em}\dp0=.36775ex\hskip0pt\box0}}
\def\ao{{}\kern-.10em\hbox{``}}

\normalsize
\begin{document}
\bibliographystyle{plain}

\begin{titlepage}
\begin{flushright}
UWThPh-1997-05
\end{flushright}
\vspace{2cm}
\begin{center}
{\Large \bf Generalized Stochastic Gauge Fixing} 
\\[40pt]
Helmuth H{\"u}ffel* and Gerald Kelnhofer** \\
Institut f{\"u}r Theoretische Physik \\
Universit\"at Wien \\
Boltzmanngasse 5, A-1090 Vienna, Austria
\vfill

{\bf Abstract}
\end{center}
We propose a generalization of the stochastic gauge fixing procedure for 
the stochastic quantization of gauge theories where not only the drift 
term of the stochastic process is changed but also the Wiener process 
itself. All gauge invariant  expectation values remain unchanged. As an 
explicit example we study the 
case of an abelian gauge field 
coupled with three bosonic matter 
fields in $0+1$ dimensions. We  
nonperturbatively  prove equivalence with the path integral formalism. 

\vfill
\begin{enumerate}
\item[*)] email: helmuth.hueffel@univie.ac.at
\item[**)] supported by ``Fonds zur F\"orderung der wissenschaftlichen 
Forschung in \"Oster\-reich",  project P10509-NAW
\end{enumerate}
\end{titlepage}

Stochastic quantization was
 presented several years ago by Parisi and Wu 
\cite{Parisi+Wu} as a novel method for the quantization of field 
theories. It provides a remarkable connection between quantum field 
theory and statistical mechanics and has grown into a useful tool in 
several areas of quantum field theory,
 see refs. \cite{Damgaard+Huffel,Namiki} for 
comprehensive reviews and referencing.

One of the  particularly interesting aspects of the stochastic 
quantization scheme lies in the quantization of 
gauge theories. Over the last years much hope has  been put forward  to 
gain new insights for a correct nonperturbative path integral
formulation 
of gauge theories also from the stochastic quantization point of view. 
However, the fundamental question  
how stochastic quantization
 -- if at all --  compares with 
the conventional quantization schemes in the case of gauge theories
  so far remained unclear and no 
really compelling argument in favour for the stochastic quatization
scheme 
 has emerged. 

In this paper we propose  a  generalized 
 stochastic gauge fixing procedure which allows to extract the 
 equilibrium Fokker-Planck probability distribution of a toy model in a 
 appealing fashion. New  hopes for related
applications also for more complicated gauge models seem justified.

The crucial point of the Parisi-Wu approach for the gauge theory 
case \cite{Parisi+Wu} is to demand that the stochastic 
time evolution of the fields is 
given  by a Langevin equation of the form
\beq
d\Phi^i(t,s) = - \left. \frac{\delta S}{\delta \Phi^i(t)}
\right|_{\Phi(t) = \Phi(t,s)} ds + dW^i
\eeq
Here we collectively denote by $\Phi^i(t,s)$, $i=1,...,m$ the pure gauge
as well as matter 
fields of the given gauge model. According to the stochastic
quantization 
procedure  these fields
 depend in addition  to their usual 
coordinates -- 
for shortness of notation  denoted by just the  single coordinate  $t$
--
on the  stochastic  
time coordinate $s$ as well. $S$ denotes the original
  (Euclidean space-time) action of 
the
given gauge 
model; it is the unmodified bare action {\it without\/} gauge symmetry 
breaking terms and {\it without\/} accompanying ghost field terms. The 
stochastic process (1) is defined in terms of the increments $dW^i$ of a 
m-dimensional Wiener
 process; it undergoes undamped diffusion 
and does not approach an equilibrium distribution. Related to this 
fact is that a Fokker-Planck formulation for the $\Phi^k$ is not 
possible because the gauge invariance of the action leads to 
divergencies in the normalization condition of the Fokker-Planck 
density \cite{Parisi+Wu}. 

Zwanziger's stochastic gauge fixing procedure \cite{Zwanziger81} 
consists in adding an additional drift force to the Langevin 
equation (1) which acts tangential to the gauge orbits. This 
additional term generally can be expressed by the components $Z^i(t',t)$ 
of the generator 
 of infinitesimal
gauge transformations and an arbitrary function $\alpha$. 
The gauge generator is given by the vector field
\beq
Z_\xi(\Phi) =
\int dt dt' \xi(t') Z^i(t',t) \frac{\delta}{\delta \Phi^i(t)} 
\eeq
 where $\xi$  
is an arbitrary element of the Lie algebra of the gauge group.
Zwanziger's modified Langevin equation reads as follows
\beq
d\Phi^i(t,s) = - \left. \left[ \frac{\delta S}{\delta \Phi^i(t)} +
\int dt'Z^i(t,t')\alpha(t')\right]
\right|_{\Phi(\cdot) = \Phi(\cdot,s)} ds + dW^i
\eeq
One can  prove that the expectation values of gauge 
invariant observables remain 
unchanged for any choice of the function $\alpha$ and 
that for specific choices of the -- in principle -- arbitrary 
function 
$\alpha$ the gauge modes' diffusion is damped along the gauge 
orbits. As a consequence the Fokker-Planck density can be 
normalized \cite{Zwanziger81}.

We present now our generalization \cite{HK} of Zwanziger's stochastic
gauge 
fixing procedure by adding a specific drift term 
which  not only has tangential  components along the gauge orbits; 
in addition we   modify the 
Wiener process itself. 
In this way we 
introduce more than just  one  function $\alpha$ , in fact we add $m$
additional  functions $\beta_i$  appearing in the 
drift term as well as in the 
Wiener process part of the Langevin equation. 

Our generalization is done in 
such a way that 
expectation values of gauge invariant observables again remain unchanged 
for any choice of $\alpha$ and $\beta_i$. The main motivation 
behind our  generalization  is that for specific 
choices of these extra  functions $\alpha$ and $\beta_i$ the  
fluctuation-dissipation theorem can be applied which  leads
to drastic simplifications 
of the stochastic process in the equilibrium limit; such a mechanism 
  is not possible 
in the original approach of Zwanziger.

Our generalized Langevin equation reads
\beqa
d\Phi^i(t,s) &=& -  \left[ \frac{\delta S}{\delta \Phi^i(t)} +
\int dt'Z^i(t,t')\alpha(t')\right. \no \\
&& \mbox{} + \left.\left. \int dt_1 dt_2
\frac{\delta Z^i(t,t_1)}{\delta \Phi^k(t_2)} \; \zeta^k(t_1,t_2)
\right] \right|_{\Phi(\cdot) = \Phi(\cdot,s)} ds \no \\
&& \mbox{} + \left. \int dt_2 \left[\delta^i{}_k \delta(t-t_2)
+ \int dt_1 Z^i(t,t_1) \beta_k(t_1,t_2)\right]
\right|_{\Phi(\cdot) = \Phi(\cdot,s)} dW^k(t_2,s) \no \\
\eeqa
We  introduced $\zeta^k(t_1,t_2)$ as a shorthand notation of
\beq
\zeta^k(t_1,t_2) = 2 \delta^{k\ell} \beta_\ell(t_1,t_2)  +
\int  dt_3 dt_4 Z^k(t_2,t_3) \beta_\ell(t_3,t_4)
\delta^{\ell m}\beta_m(t_1,t_4). 
\eeq
We see that the new drift term clearly is not acting tangential to 
the gauge orbit; its rather complicated structure is necessary for 
leaving unchanged gauge invariant expectation values; the
straightforward 
proof is 
given in \cite{HK}. 

In order to proceed more explicitely we  decided to study the so called 
Helix model, which 
describes the minimal coupling of an abelian gauge field with 
three bosonic matter fields in $0+1$ dimensions. This model was 
originally proposed by deWit \cite{deWit}  and was investigated 
intensively within the Hamiltonian framework by Kucha$\check{\rm r}$ 
\cite{Kuchar}. Recently the helix model came to new life again 
\cite{Friedberg,Fujikawa} in the course of studies on problems with
gauge 
fixing.
The helix model is defined by the Lagrange density
\beqa
L(t) = \frac{1}{2} [(\dot \vp^1 - A \vp^2)^2 
  \mbox{}+(\dot \vp^2 + A \vp^1)^2 
+(\dot \vp^3 - A)^2] 
 \mbox{}-\frac{1}{2} [(\vp^1)^2 + (\vp^2)^2]
\eeqa
where the dot denotes time derivation and the fields
$(\Un{\vp}(t),\vp^3(t)) = (\vp^1(t),\vp^2(t),\vp^3(t))$
and $A(t)$ are regarded as elements of the function spaces 
$\E = C^\infty({\bf R},{\bf R}^3)$ and $\A = C^\infty({\bf R},{\bf R})$,
respectively. Hence the total number of gauge and matter fields is $m=4$
and
$\Phi=(\Un{\vp},\vp^3,A)$.
Let $\G = C^\infty({\bf R},{\bf R})$ denote the abelian group of 
gauge transformations and consider the following transformation 
on the configuration space
\beq
(\Un{\vp},\vp^3,A) \rightarrow  (R(g) \Un{\vp}, \vp^3 - g,A - 
\dot g)
\eeq
where $g \in \G$ and
$$
R(g) = \left( \ba{rr} 
\cos g & - \sin g \\ \sin g & \cos g \ea \right).
$$
The Lagrange density $L(t)$ is easily verified to be invariant under
these
transformations. The components $Z^i(t',t)$ of the generator of
infinitesimal
gauge transformations can be read off the vector field
\beqa
Z_\xi(\Phi) 
&=& \int _{\bf R} dt (- \xi \vp^2 
\frac{\delta}{\delta \vp^1(t)} + \xi \vp^1\frac{\delta}
{\delta \vp^2(t)}  
- \xi \frac{\delta}{\delta \vp^3(t)} -
\dot \xi \frac{\delta}{\delta A(t)} )
\eeqa

We rewrite the stochastic process in 
terms of gauge invariant and gauge dependent fields (for the explicit 
geometrical structure see \cite{HK})
\beqa
\Un{\Psi} &=& R(\vp^3) \; \Un{\vp} \no \\
\Psi^3 &=& A - \dot \vp^3 \\
\Psi^4 &=&  - \vp^3 \no
\eeqa
In these new coordinates, gauge transformations are given purely as
translations, i.e. 
$(\Un{\Psi},\Psi^3,\Psi^4) \ra
(\Un{\Psi},\Psi^3,\Psi^4 - g)$
where $g \in \G$.
With respect to these 
variable changes we introduce the vielbeins $E$ and their inverses 
$e$
\beq
E^\mu{}_i(t,t') = \frac{\delta \Psi^\mu(t)}{\delta \Phi^i(t')}, \qquad
e^i{}_\mu(t,t') = \frac{\delta \Phi^i(t)}{\delta \Psi^\mu(t')}, 
\eeq
as well as the induced inverse metric $G^{\mu\nu}$ 
\beq
G^{\mu\nu}(t_1,t_2) = \int_{\bf R} dt_3 \; E^\mu{}_i(t_1,t_3)
\delta^{ij} E_j{}^\nu(t_2,t_3).
\eeq
We can choose 
such specific values for the functions $\alpha$ and $\beta_k$  
that the gauge modes' diffusion is 
damped along the gauge 
orbits (as a consequence the Fokker-Planck density can be 
normalized) and that  the equilibrium limit of the 
stochastic process can explicitely be derived. We take
\beq
\alpha(t) = \int_{\bf R} dt' [ G^{4\nu}(t,t')
\frac{\delta S}{\delta \Psi^\nu(t')} -
\gamma(t,t') \Psi^4(t')]
\eeq
as well as 
\beq
\beta_k(t_1,t_2) = - E^4{}_k(t_1,t_2) + \delta_{k\ell} e^\ell{}_4
(t_1,t_2)
\eeq
and
\beq
\gamma(t_1,t_2) = \int_{\bf R} dt_3 e^k{}_4(t_1,t_3) \delta_{k\ell}
e^\ell{}_4(t_2,t_3)
\eeq
The Langevin equations now read (according to the rules of Ito's
stochastic 
calculus)
\beqa
d\Psi^\mu(t,s) &=& \int_{\bf R} dt_1 \left\{ \left[ -\wt G^{\mu\nu}
(t_,t_1) \frac{\delta S_{\rm tot}}{\delta \Psi^\nu(t_1)} +
\frac{\delta \wt G^{\nu\mu}(t,t_1)}{\delta \Psi^\nu(t_1)} \right]ds
\right. \no \\
&& \mbox{} + \left.\left. \wt E^\mu{}_k(t,t_1) dW^k(t_1,s)\right\}
\right|_{\Psi(\cdot) = \Psi(\cdot,s)}
\eeqa
Here we introduced the 
total action $S_{\rm tot}$ 
\beq
S_{\rm tot} = S + \frac{1}{2} \int_{\bf R}  dt
(\Psi^4(t))^2,
\eeq
 the matrix $\wt G$ 
\beqa
\wt G^{\bar\mu \bar\nu}(t_1,t_2) &=& G^{\bar\mu \bar\nu}(t_1,t_2)
\qquad \bar\mu, \bar\nu = 1,2,3 \no \\
\wt G^{\bar\mu 4}(t_1,t_2) &=& \wt G^{4 \bar\mu}(t_1,t_2) = 0 \no \\
\wt G^{44}(t_1,t_2) &=& \gamma(t_1,t_2)
\eeqa
and the vielbein $\wt E$
\beqa
\wt E^\mu{}_k(t_1,t_2) = E^\mu{}_k(t_1,t_2) + \delta^\mu{}_4 
[- E^4{}_k(t_1,t_2) + \delta_{k\ell} e^\ell{}_4
(t_1,t_2)]
\eeqa
We remark that $\wt G$ is explicitly decomposable as
\beq
\wt G^{\mu\nu}(t_1,t_2) = \int_{\bf R} dt_3 \wt E^\mu{}_k(t_1,t_3)
\delta^{k\ell} \wt E^\nu{}_\ell(t_2,t_3)
\eeq
and by construction is a positive matrix. 

We now derive the equilibrium distribution of the stochastic 
process described by the above Langevin equation (15) by studying the 
corresponding  
Fokker-Planck equation. We remind that we restricted ourselves to 
a well converging stochastic processes, so 
that the Fokker-Planck probability distribution is normalizable. 
Most crucially we have that $\wt G$ is positive and is 
appearing in the Fokker-Planck operator in factorized form.
As a consequence  (due to the fluctuation-dissipation theorem)
 the formal stationary limit of the Fokker-
Planck probability distribution can be identified with the 
equilibrium limit and reads
\beq
\rho[\Psi]_{\rm equil.} = \frac{e^{-S_{\rm tot}}} {\int D \Psi
e^{- S_{\rm tot}}}.
\eeq
After integrating out $\Psi^3$ and $\Psi^4$ \cite{HK}, our path 
integral density  is agreeing nicely with the result of
\cite{Friedberg}.

\end{document}